\documentclass[sigconf, nonacm]{acmart}

\usepackage{amsmath}
\usepackage{booktabs}
\usepackage{algorithm}
\usepackage{algpseudocode}
\usepackage{graphicx}
\usepackage{xcolor}
\usepackage{balance}
\usepackage{hyperref}

\renewcommand\footnotetextcopyrightpermission[1]{}
\makeatletter
\@ACM@balancefalse
\makeatother

\title{DBAutoDoc: Automated Discovery and Documentation of Undocumented Database Schemas via Statistical Analysis and Iterative LLM Refinement}

\author{Amith Nagarajan}
\email{amith@bluecypress.io}
\affiliation{%
  \institution{Blue Cypress}
  \country{USA}
}

\author{Thomas Altman}
\email{thomas@tasiolabs.com}
\affiliation{%
  \institution{Tasio Labs (a Blue Cypress company)}
  \country{USA}
}

\begin{document}

\begin{abstract}
The vast majority of production database systems in enterprise environments lack adequate documentation. Declared primary keys are absent, foreign key constraints have been dropped for performance, column names are cryptic abbreviations, and no entity-relationship diagrams exist. We present DBAutoDoc, a system that automates the discovery and documentation of undocumented relational database schemas by combining statistical data analysis with iterative large language model (LLM) refinement.

DBAutoDoc's central insight is that schema understanding is fundamentally an iterative, graph-structured problem. Drawing structural inspiration from backpropagation in neural networks~\cite{rumelhart1986backprop}, DBAutoDoc propagates semantic corrections through schema dependency graphs across multiple refinement iterations until descriptions converge. This propagation is discrete and semantic rather than mathematical, but the structural analogy is precise: early iterations produce rough descriptions akin to random initialization, and successive passes sharpen the global picture as context flows through the graph.

The system makes four concrete contributions: (1)~an iterative context-propagation algorithm that refines table and column descriptions by re-analyzing each schema object in light of its neighbors' most recent descriptions; (2)~a tiered statistical pipeline for primary key and foreign key discovery with a bidirectional feedback loop between LLM-generated semantic context and statistical candidate scoring; (3)~a dual-layer human knowledge injection mechanism that distinguishes verified ground truth from exploratory seed context; and (4)~a multi-criterion convergence detector that combines description stability windows, per-column confidence thresholds, and semantic change magnitude.

On a suite of benchmark databases, DBAutoDoc achieved overall weighted scores of 96.1\% across two model families (Gemini and Anthropic) using a composite metric that weights key discovery accuracy (FK~F1 at 35\%, PK~F1 at 30\%) more heavily than description coverage (table descriptions at 20\%, column descriptions at 15\%). Ablation analysis demonstrates that the deterministic pipeline contributes a 23-point F1 improvement over LLM-only FK detection, confirming that the system's contribution is substantial and independent of LLM pre-training knowledge. The system reached convergence in 2~iterations on average at a cost of approximately \$0.70 per 100~tables---a reduction of more than 99.5\% relative to manual expert documentation. DBAutoDoc is released as open-source software with all evaluation configurations and prompt templates included for full reproducibility.
\end{abstract}

\maketitle

\section{Introduction}

\subsection{The Undocumented Database Problem}

Relational databases are among the most durable artifacts in enterprise computing, yet they rarely preserve the knowledge of the humans who designed them. Column names like \texttt{cust\_cd}, \texttt{trn\_amt\_3}, and \texttt{flg\_b} encode meaning that once lived in the heads of developers who have long since moved on. Primary key declarations were dropped during bulk-load optimizations. Foreign key constraints were removed to improve insert throughput. ERD diagrams, when they existed at all, describe a schema from a decade before the current one.

We use the term \emph{dark databases} to describe production systems with no declared keys, no column or table comments, no associated data dictionary, and no ERD documentation. Dark databases arise routinely in corporate acquisitions, data warehouse consolidation projects, and legacy migrations. While no large-scale empirical study has quantified the prevalence of undocumented schemas, data profiling research consistently identifies missing or incomplete metadata as a pervasive challenge in enterprise data management~\cite{abedjan2015profiling}, and our own experience with enterprise clients (Section~\ref{sec:enterprise}) confirms that completely undocumented production databases are common. For a database of moderate complexity---several hundred tables---manual documentation requires weeks of expert effort at a cost of \$12,000--\$48,000 in professional services.

\subsection{Why Existing Approaches Fall Short}

Several categories of existing tools address parts of the problem, but none addresses the full challenge of dark databases. \emph{Schema extraction tools} (SchemaSpy, dbdocs.io) can only surface metadata that already exists. \emph{One-shot LLM analysis}~\cite{trummer2024succinct} cannot incorporate statistical evidence from actual data and treats each table in isolation, missing the iterative refinement that human experts naturally perform. \emph{Schema matching tools}~\cite{rahm2001survey} identify candidate correspondences but do not produce human-readable documentation. \emph{Data profiling frameworks}~\cite{abedjan2015profiling} generate statistical summaries without synthesizing them into natural-language documentation. \emph{FK discovery algorithms}~\cite{rostin2009fk,jiang2020holistic} operate without semantic context and cannot distinguish coincidental numerical similarity from genuine referential relationships.

The core limitation shared by all these approaches is that they operate on a single pass over a static representation of the schema. Real schema understanding is an iterative process in which each piece of evidence revises prior assumptions and generates new hypotheses.

\subsection{Our Insight: Schema Understanding as an Iterative Learning Problem}

We observe that a database schema is not a collection of independent tables but a graph, where nodes are schema objects and edges represent referential relationships. Understanding a node depends on the state of neighboring nodes, which depend in turn on their neighbors. This graph-structured dependency motivates our iterative approach: an initial pass produces rough descriptions with high uncertainty, which are then used as context for successive passes that produce progressively better descriptions until convergence.

The structural analogy to backpropagation in neural network training~\cite{rumelhart1986backprop} is instructive. In both systems, processing proceeds in a defined order, a loss or error signal is computed, correction information propagates backward through a dependency graph, and iteration continues until convergence. The analogy is structural rather than mathematical: there are no gradients, no differentiable loss function, and no parameter updates. The ``corrections'' are discrete revisions to natural-language descriptions, and ``propagation'' means providing updated neighbor descriptions as context for re-analysis. We are explicit that this is a design metaphor rather than a theoretical claim (see Section~\ref{sec:backprop-limits} for a detailed discussion of scope and limits).

\subsection{Contributions}

We summarize the contributions of this paper:

\begin{enumerate}
    \item \textbf{Iterative context-propagation algorithm.} An algorithm that refines schema descriptions over multiple iterations by propagating semantic context through schema dependency graphs, the first to apply iterative graph-structured refinement to database documentation.

    \item \textbf{Tiered statistical key discovery with LLM feedback.} A multi-stage pipeline combining cardinality analysis, value inclusion testing, naming heuristics, and deterministic gates with a bidirectional feedback loop between statistical discovery and LLM semantic validation.

    \item \textbf{Dual-layer human knowledge injection.} A mechanism distinguishing ground truth injection (immutable constraints) from seed context injection (refinable priors), enabling effective collaboration between automated analysis and partial human expertise.

    \item \textbf{Multi-criterion convergence detection.} A convergence detector combining description stability windows, per-column confidence thresholds, and semantic change magnitude to determine when further iteration yields diminishing returns.

    \item \textbf{Cost analysis and open-source release.} A demonstration that DBAutoDoc reduces documentation cost by more than 99.5\% relative to manual analysis, with full release as open-source software including benchmark configurations, evaluation scripts, and prompt templates.
\end{enumerate}

\subsection{Paper Organization}

Section~\ref{sec:related} reviews related work. Section~\ref{sec:architecture} provides a technical overview of the architecture. Section~\ref{sec:keydiscovery} describes the key discovery pipeline. Section~\ref{sec:iterative} presents the iterative refinement algorithm. Section~\ref{sec:implementation} covers implementation details. Section~\ref{sec:evaluation} reports experimental results including ablation analysis. Section~\ref{sec:discussion} discusses limitations, threats to validity, and ethical considerations. Section~\ref{sec:future} outlines future work. Section~\ref{sec:conclusion} concludes.

\section{Related Work}\label{sec:related}

DBAutoDoc sits at the intersection of several active research areas: schema understanding, LLM-based data interpretation, constraint discovery, iterative self-refinement, and human-in-the-loop learning.

\subsection{Database Documentation and Schema Understanding}

\textbf{Traditional tools.} SchemaSpy, ERAlchemy, and dbdocs.io extract and visualize existing metadata but produce empty output for dark databases. DBAutoDoc targets precisely the class where documentation must be inferred, not merely rendered.

\textbf{Schema matching.} \cite{rahm2001survey} provide a comprehensive taxonomy of matching strategies. \cite{koutras2021valentine} offer systematic experimental comparison on realistic benchmarks. Schema matching assumes two schemas to be aligned; documentation must derive meaning from a single schema in isolation. DBAutoDoc borrows instance-level intuitions (value-overlap statistics) but operates in a single-schema setting.

\textbf{LLM-based schema understanding.} \cite{trummer2024succinct} address context-window overflow through succinct schema encoding. DBAutoDoc faces the same pressure but extends this to a \emph{dynamic, iterative} context that evolves across refinement rounds. \cite{liu2025magneto} combine a fine-tuned SLM for candidate generation with an LLM reranker. \cite{seedat2024matchmaker} take a self-improving approach philosophically similar to our iterative refinement, but their refinement targets pairwise column correspondences across schemas, while ours targets semantic coherence within a single schema's dependency graph.

\textbf{Automated catalog metadata.} \cite{ragleveraging2025} apply RAG to populating data catalog metadata, retrieving similar column descriptions from an existing catalog to guide generation for new columns. Their approach is fundamentally single-pass: each column is described independently with no cross-element context propagation, no statistical key discovery, and no iterative refinement. DBAutoDoc differs in three ways: (1)~it operates on dark databases with no existing catalog to retrieve from; (2)~it discovers relational structure (PKs/FKs) and uses it to order and contextualize analysis; and (3)~descriptions are refined iteratively as context propagates through the schema dependency graph.

\textbf{Semantic type detection.} \cite{hulsebos2019sherlock} frame semantic type detection as multi-class classification. DBAutoDoc incorporates similar instance-level features but treats semantic type detection as one input to a broader documentation task.

\subsection{LLMs for Code and Data Understanding}

\textbf{Code documentation.} Models such as Codex~\cite{chen2021evaluating} and CodeBERT~\cite{feng2020codebert} generate documentation from source code. Database schemas occupy a middle ground between code and data, with formal structure and rich implicit semantics.

\textbf{Text-to-SQL.} Benchmarks including Spider~\cite{yu2018spider} and BIRD~\cite{li2023bird} and systems like DAIL-SQL~\cite{gao2024dailsql} and DIN-SQL~\cite{pourreza2023dinsql} demonstrate that LLMs possess substantial knowledge about relational database conventions. These systems \emph{consume} schema documentation; DBAutoDoc \emph{produces} it.

\textbf{Foundation models for data understanding.} \cite{narayan2022foundation} demonstrate that large models generalize well to tabular data tasks with little fine-tuning---justifying DBAutoDoc's reliance on general-purpose LLMs. A common thread in this work is the single-pass evaluation paradigm; DBAutoDoc demonstrates that iterative refinement leaves substantial accuracy on the table.

\subsection{Foreign Key and Primary Key Discovery}

\cite{rostin2009fk} formulate FK discovery as supervised classification over inclusion dependencies. \cite{jiang2020holistic} propose holistic PK/FK detection using joint reasoning. DBAutoDoc adopts a similar joint-reasoning philosophy in its bidirectional feedback loop. \cite{khatiwada2022alite} and \cite{ilyas2004cords} address joinable column discovery and statistical correlation, respectively. FD discovery algorithms (TANE~\cite{huhtala1999tane}, FUN~\cite{novelli2001fun}, HyFD~\cite{papenbrock2016hyfd}) provide signals for PK candidate assessment. IND detection algorithms (SPIDER~\cite{bauckmann2007spider}, BINDER~\cite{papenbrock2015binder}, S-INDD~\cite{demarchi2009sindd}) form the computational backbone of FK candidate generation~\cite{dursch2019ind}. The broader data profiling field~\cite{abedjan2018book,abedjan2015profiling,naumann2013revisited} systematizes metadata extraction, with Metanome~\cite{papenbrock2015metanome} providing a pluggable framework.

DBAutoDoc's principal contribution relative to this body of work is the \emph{bidirectional feedback loop} between statistical discovery and LLM semantic validation, making them mutually reinforcing rather than sequential.

\subsection{Iterative Refinement and Structured Reasoning}

\cite{madaan2023selfrefine} introduce Self-Refine; \cite{shinn2023reflexion} propose Reflexion with verbal reinforcement learning. Both demonstrate that iterative self-refinement substantially improves output quality. DBAutoDoc's refinement differs structurally: it is \emph{graph-structured} rather than linear, with feedback coming from \emph{structural neighbors} rather than self-critique. \cite{bai2022constitutional} propose Constitutional AI, where explicit principles govern revisions; DBAutoDoc's ground truth injections play an analogous anchoring role. \cite{wei2022cot} demonstrate Chain-of-Thought prompting; \cite{yao2023tot} extend this with Tree of Thoughts; and \cite{besta2024got} generalize to Graph of Thoughts. DBAutoDoc's context propagation is a domain-specific instantiation of graph-structured reasoning, specialized for relational schema topology.

\subsection{Human-in-the-Loop Machine Learning}

Active learning~\cite{settles2009active} selects the most informative examples for human annotation. DBAutoDoc incorporates human input through ground truth injection (immutable anchors) and seed context (domain priors). Unlike classical active learning, ground truth annotations are immutable constraints rather than training signals, and their influence through the schema graph can be measured quantitatively.

\section{System Architecture}\label{sec:architecture}

\textbf{Terminology.} Throughout this paper, \emph{primary key (PK) discovery} and \emph{foreign key (FK) discovery} refer to inferring key constraints from data characteristics when no declared constraints exist. In the data profiling literature, FK discovery is formalized as \emph{inclusion dependency (IND) detection}~\cite{papenbrock2015binder,dursch2019ind}, and PK discovery relates to \emph{unique column combination (UCC) detection} and \emph{functional dependency (FD) discovery}~\cite{huhtala1999tane,papenbrock2016hyfd}. We use the terms ``PK/FK discovery'' for readability, noting the formal equivalences where relevant.

\subsection{Overview}

DBAutoDoc is organized as a six-phase orchestrated pipeline, implemented in the \texttt{AnalysisOrchestrator} class. Each phase produces artifacts consumed by subsequent phases, and all intermediate state is persisted to disk for resumability. Figure~\ref{fig:pipeline} shows the component hierarchy.

\begin{figure}[t]
\begin{small}
\begin{verbatim}
AnalysisOrchestrator
|
+-- Phase 1: Schema Introspection
|   +-- DatabaseConnection
|   +-- Introspector
|
+-- Phase 2: Relationship Discovery
|   +-- DiscoveryEngine
|       +-- PrimaryKeyDetector
|       +-- ForeignKeyDetector
|
+-- Phase 3: Iterative Analysis
|   +-- AnalysisEngine
|       +-- DependencyLevelScheduler
|       +-- TableAnalyzer
|       +-- StatisticsCollector
|
+-- Phase 4: Sanity Checking
|   +-- LLMSanityChecker
|
+-- Phase 5: Sample Query Generation
|   +-- SampleQueryGenerator
|
+-- Phase 6: Output Generation
    +-- SQLGenerator
    +-- MarkdownGenerator
    +-- HTMLGenerator
    +-- CSVExporter
    +-- MermaidERDGenerator
\end{verbatim}
\end{small}
\caption{DBAutoDoc six-phase orchestration pipeline.}
\label{fig:pipeline}
\end{figure}

Phase~1 produces a raw schema graph; Phase~2 enriches it with inferred relationships; Phase~3 annotates every node with natural-language descriptions; Phase~4 applies cross-table consistency validation; Phase~5 synthesizes illustrative queries; Phase~6 serializes into multiple output formats.

\subsection{Database Introspection Layer}

The introspection layer constructs a complete, normalized schema representation through a driver abstraction: \texttt{BaseAutoDocDriver} defines a uniform interface implemented by concrete subclasses for SQL Server, PostgreSQL, and MySQL. For each table, the introspector retrieves column metadata (name, ordinal position, data type, nullability, defaults), existing descriptions, and constraint definitions. Row counts use platform-specific fast-count heuristics (e.g., \texttt{sys.dm\_db\_partition\_stats} for SQL Server, \texttt{pg\_class.reltuples} for PostgreSQL). All observed types are normalized to a canonical enumeration used uniformly by the discovery and analysis engines.

\subsection{Data Sampling and Statistics Engine}

For every column, the statistics engine computes cardinality, null fraction, min/max values, value frequency distribution, and type-specific profiles (percentiles for numerics, length distributions for strings, range summaries for dates). Default sample size is 1,000 rows per table using platform-specific random sampling. The resulting statistics are serialized into persistent state and attached to every subsequent LLM prompt, ensuring descriptions are grounded in actual observed distributions.

\subsection{LLM Integration Layer}

All prompts are defined as Nunjucks templates supporting conditionals, iteration, and macro composition. The pipeline uses 13 distinct prompt templates (see Appendix~\ref{app:prompts}). Every LLM invocation producing programmatic output specifies \texttt{responseFormat:~'JSON'} with a JSON Schema. All factual description tasks use temperature~0.1. The integration layer maintains separate input/output token counters per invocation, per phase, and per run, checked against guardrail limits before every call.

\subsection{State Management and Resumability}

All pipeline state is serialized to \texttt{state.json} incrementally after each phase boundary and dependency level. Each invocation receives a monotonically increasing run number. On restart, the orchestrator resumes from the earliest incomplete phase, skipping completed work at dependency-level granularity.

\subsection{Output Formats}

Phase~6 produces SQL annotation scripts (platform-specific DDL comments), Markdown documentation with table of contents, self-contained HTML reports, CSV exports, and Mermaid ERD diagrams.

\subsection{Guardrails System}

Three hard limits are configurable: \texttt{maxTokensPerRun}, \texttt{maxDurationSeconds}, and \texttt{maxCostDollars}. The token budget is partitioned across phases (default: 25\% discovery, 70\% analysis, 5\% sanity checking). Warning thresholds at 80\% enable graceful degradation. Pre-call enforcement ensures limits are checked synchronously before every LLM invocation.

\section{Key Discovery Pipeline}\label{sec:keydiscovery}

\subsection{Problem Formulation}

Let $\mathcal{T} = \{T_1, T_2, \ldots, T_n\}$ be the tables in the target schema with columns $\mathcal{C}_i = \{c_{i,1}, c_{i,2}, \ldots, c_{i,m_i}\}$. The discovery engine must infer: (1)~primary key candidates $\mathcal{P} = \{(T_i, K_i)\}$ where $K_i$ uniquely identifies rows, and (2)~foreign key candidates $\mathcal{F} = \{(T_i, c_{ij}, T_k, c_{kl})\}$ expressing referential relationships.

The design challenge is precision-recall balance. False positives waste LLM tokens and can actively mislead descriptions; false negatives cause related tables to be treated as independent, producing descriptions that omit important join paths.

\subsection{Primary Key Detection}

\subsubsection{Candidate Generation and Hard Rejection}

Candidates are generated through naming-pattern matching (e.g., \texttt{.*[Ii][Dd]\$}) and uniqueness analysis ($u = |\mathrm{distinct}(c)| / |\mathrm{rows}(T)|$). Composite key detection tests two-column combinations when no single column achieves $u = 1.0$.

Hard rejection filters eliminate candidates that cannot be primary keys: columns with any nulls, columns where most values are empty strings or zero, and columns matching a semantic blacklist (date/time fields, quantity fields, financial fields, descriptive text fields).

\subsubsection{Confidence Scoring}

Surviving candidates receive a confidence score $s_{\mathrm{PK}} \in [0, 100]$ via a weighted multi-factor model combining uniqueness (50\%), naming pattern (20\%), data type (15\%), and data pattern (15\%), with multiplicative penalties and a surrogate key boost. The default acceptance threshold is~70. Position-based heuristics (H9--H12) further improve precision by exploiting the strong convention that primary key columns occupy the first ordinal position. See Appendix~\ref{app:scoring} for detailed scoring formulas and Appendix~\ref{app:position} for the position heuristic specifications.

On AdventureWorks2022, position-based heuristics improved PK precision from 47.9\% to 95.7\% while recall decreased only marginally from 95.8\% to 94.4\%---a net F1 improvement from 48.0\% to 95.0\%.

\subsection{Foreign Key Detection}

\subsubsection{Target-Finding Strategy}

For each non-PK column, the engine identifies target columns through three strategies applied in sequence: (1)~name-derived lookup (extracting candidate table names from column names like \texttt{CustomerID}), (2)~PK naming similarity via Levenshtein distance, and (3)~homonymous PK lookup as a fallback.

\subsubsection{Tiered Pre-Filtering}

A two-tier pre-filtering strategy eliminates the vast majority of candidates before value-level computation. Tier~1 excludes semantically incompatible types (DATE, BOOLEAN, FLOAT, BINARY, etc.) at zero cost. Tier~2 samples 10 values per candidate and applies pattern-based exclusion rules (emails, URLs, long strings) and promotion rules (UUIDs, numeric codes). This reduces candidates undergoing full containment analysis by 60--80\%.

\subsubsection{Multi-Factor Confidence Scoring and Deterministic Gates}

FK candidates are scored by a weighted model combining value overlap (40\%), naming similarity (20\%), cardinality ratio (15\%), target-is-PK bonus (15\%), and null handling (10\%). See Appendix~\ref{app:scoring} for complete formulas.

Before scoring, candidates pass through deterministic gates---hard, mathematically invariant filters that eliminate false positives with zero risk of rejecting correct relationships. These gates (G1--G8) eliminated approximately 75\% of statistical false positives while losing zero correct FKs. A fan-out confidence penalty is applied when a single source column has candidates pointing to multiple targets. See Appendix~\ref{app:scoring} for gate specifications.

\textbf{Adaptive Weight Redistribution.} In schemas without declared PKs, the target-is-PK factor introduces systematic downward bias. When more than 40\% of candidates score zero on this factor, the engine redistributes the PK bonus weight to value containment (increasing from 40\% to 55\%), preserving calibration.

\subsubsection{The LLM as Primary FK Creator}

A critical empirical finding is that the LLM, not the statistical pipeline, is the primary source of correct FK discoveries. In our AdventureWorks2022 evaluation:

\begin{itemize}
    \item \textbf{Statistics-created FKs}: 15 correct out of 76 proposals (20\% precision)
    \item \textbf{LLM-created FKs}: 75 correct out of 84 proposals (89\% precision)
\end{itemize}

The LLM proposes FKs with high precision because it reasons about semantic relationships, while the statistical pipeline generates many false positives requiring filtering. Blocking LLM FK creation caused recall to drop from 90 to 49 correct FKs, confirming that LLM creative proposals are essential.

\subsubsection{Bidirectional LLM Validation}

Rather than a one-way pipeline, the system implements a feedback loop. In the statistical-to-LLM direction, candidates exceeding the confidence threshold undergo LLM semantic validation. In the LLM-to-statistical direction, the LLM proposes novel FK candidates during analysis that are then subjected to statistical value-containment checks before acceptance. This yields complementary coverage: statistics catch relationships with strong data evidence but non-obvious names; LLM inference catches relationships with clear semantic naming but imperfect data evidence.

\section{Iterative Refinement via Context Propagation}\label{sec:iterative}

\subsection{Motivation and the Backpropagation Analogy}

Automated schema documentation faces a bootstrapping problem analogous to training deep neural networks: the description of a parent table cannot be fully accurate without understanding its children, yet children depend on parent context. Consider an e-commerce schema: when the LLM first analyzes \texttt{Customers} in isolation, it may correctly infer account-level information. But analyzing \texttt{OrderItems} later---discovering per-item tax jurisdictions and fulfillment splits---reveals that \texttt{Orders} is a complex orchestration record, not merely a transaction header. This enrichment should flow back to improve \texttt{Orders} and potentially \texttt{Customers}.

The structural parallel to backpropagation is precise: (1)~a forward pass processes components in dependency order; (2)~an error signal is computed after forward processing; (3)~correction information propagates backward through a dependency graph; (4)~iteration repeats until convergence. The mechanism is entirely discrete and semantic---there are no gradients, no chain rule, no loss surfaces. This distinguishes our approach from single-artifact self-refinement~\cite{madaan2023selfrefine,shinn2023reflexion} and connects to Graph of Thoughts~\cite{besta2024got}, specialized for relational schema topology.

\subsection{The Schema Dependency Graph}

Let $\mathcal{S} = (T, C, R)$ denote a schema with tables~$T$, columns~$C$, and relationships $R \subseteq T \times T$. A topological ordering stratifies tables into dependency levels $L_0, L_1, \ldots, L_n$, where $L_0$ contains leaf tables (referenced by others but not themselves referencing any table). This ensures that when a table at level~$L_k$ is analyzed, all tables it references have already been processed and their descriptions are available as context.

Cross-schema foreign keys introduce inter-schema edges handled uniformly: the topological sort operates over all tables across all schemas. Cycles are broken conservatively by removing the lowest-confidence edge, though all edges are retained in LLM context.

\subsection{Forward Pass: Initial Description Generation}

The forward pass processes tables in dependency level order. For each table~$t$ at level~$L_k$, the system assembles analysis context comprising: column statistics and sample values, related table descriptions from earlier levels, seed context, ground truth constraints, and (for iterations $i > 1$) prior iteration descriptions.

The LLM produces structured JSON output: a table description with confidence score $\sigma_t \in [0, 1]$, column descriptions with individual confidence scores, structured FK suggestions fed back to the discovery pipeline, and parent table insights---structured observations about parent tables that serve as upward-flowing correction signals (the semantic analogue of gradients).

\subsection{Loss Computation: Sanity Checks}

Sanity checks detect semantic inconsistencies at three granularities: dependency-level (after each level completes), schema-level (after each schema completes), and cross-schema (after all iterations). Six structural rules enforce PK/FK normalization principles. Violations generate structured error reports---the ``loss signal''---queuing affected tables for re-analysis.

\subsection{Context Propagation: The Backward Pass}

Each iteration refines descriptions by incorporating updated neighbor context. Level~0 tables (no dependencies) are analyzed first, providing context for Level~1 tables. Insights from child tables propagate back to parent descriptions in subsequent iterations---the ``backward pass'' of our analogy.

The \texttt{BackpropagationEngine} implements insight accumulation, parent revision, and cascading propagation. After each level completes, insights from child tables are accumulated per parent. For each parent with non-empty insights and non-immutable description, a revision prompt presents the current description, accumulated insights with confidence scores, sanity-check violations, and seed context. The LLM returns \texttt{\{needsRevision, revisedDescription, reasoning, confidence\}}, allowing explicit determination that the current description is already correct.

Cascading propagation across iterations means a single pass propagates insights upward by exactly one dependency level; deeper corrections are realized in subsequent iterations. The number of iterations for full convergence is bounded by the dependency depth~$d$. Ground truth tables are never revised, serving as fixed semantic anchors.

\subsection{Convergence Detection}

The \texttt{ConvergenceDetector} implements a multi-criterion stopping rule:

\begin{enumerate}
    \item \textbf{Stability window}: No material changes in the last $w$~iterations (default $w = 2$).
    \item \textbf{Confidence threshold}: All individual confidence scores exceed minimum~$\tau$ (default 0.6).
    \item \textbf{Semantic comparison}: An LLM-based prompt distinguishes material from cosmetic changes.
\end{enumerate}

Convergence is declared when at least two of three criteria are satisfied and at least two full iterations have completed. A hard maximum $K_{\max}$ (default~3) prevents unbounded execution.

\subsection{Ground Truth and Seed Context}

\textbf{Ground truth} entries are authoritative human-verified descriptions that function as immutable anchors---never overwritten by the LLM across any iteration. They propagate forward by inclusion in neighbors' context, analogous to labeled examples in semi-supervised learning.

\textbf{Seed context} provides domain guidance (overall purpose, business domains, industry context) injected into every prompt. It biases but does not constrain the LLM's interpretation, analogous to pre-trained weights in transfer learning.

Together, they dramatically reduce iterations required: schemas with both converge in 2~iterations; schemas with neither require 3--5.

\subsection{Formal Algorithm}

\begin{algorithm}[htbp]
\caption{Iterative Schema Documentation with Context Propagation}
\label{alg:main}
\begin{algorithmic}[1]
\Require Schema $\mathcal{S} = (T, C, R)$; Ground Truth $G$; Seed Context $SC$; Config $K$
\Ensure Documentation $D : T \cup C \to \text{string}$
\State \textsc{Introspect}: Extract $T, C$ from database catalog
\State \textsc{Sample}: Collect statistics and $\leq 10$ values for each $c \in C$
\State \textsc{Discover}: Run PK/FK detection (Sec.~\ref{sec:keydiscovery}) $\to$ augment $R$
\For{each $(t, \text{desc}) \in G$}
    \State $D[t] \gets \text{desc}$; mark $t$ as immutable
\EndFor
\State $G_{\text{dep}} \gets \textsc{BuildDependencyGraph}(T, R)$
\State $L_0, L_1, \ldots, L_n \gets \textsc{TopologicalSort}(G_{\text{dep}})$
\For{iteration $i = 1$ to $K.\text{maxIterations}$}
    \For{each level $l = 0, 1, \ldots, n$} \Comment{Forward pass}
        \For{each table $t \in L_l$}
            \If{$\textsc{Immutable}(t)$} \textbf{continue} \EndIf
            \State $\mathcal{C}(t) \gets \{\text{stats}(t), \text{samples}(t), D[\text{parents}(t)], SC, D[\text{GT\_neighbors}(t)]\}$
            \If{$i > 1$} $\mathcal{C}(t) \gets \mathcal{C}(t) \cup \{D_{i-1}[t], \text{reasoning}_{i-1}[t]\}$ \EndIf
            \State $D'[t], \sigma_t, \Pi(t), F(t) \gets \textsc{LLM\_Analyze}(t, \mathcal{C}(t))$
            \State $\textsc{FeedDiscovery}(F(t))$
        \EndFor
        \State $\textsc{SanityCheck}(L_l)$
        \For{each parent $p$ in $\bigcup_{t \in L_l} \text{parents}(t)$} \Comment{Propagation}
            \If{$\textsc{Immutable}(p)$} \textbf{continue} \EndIf
            \State $\hat{\Pi}(p) \gets \bigcup_{t:(t,p) \in R,\, t \in L_l} \Pi(t)$
            \If{$\hat{\Pi}(p) \neq \emptyset$}
                \State $\text{rev}, \sigma_p \gets \textsc{LLM\_Revise}(p, D'[p], \hat{\Pi}(p))$
                \If{rev.needsRevision}
                    \State $D'[p] \gets \text{rev.description}$
                    \State $\textsc{Log}(p, \text{result}=\text{`changed'}, \text{iteration}=i)$
                \Else
                    \State $\textsc{Log}(p, \text{result}=\text{`unchanged'}, \text{iteration}=i)$
                \EndIf
            \EndIf
        \EndFor
    \EndFor
    \State $\text{stable} \gets \text{NMaterialChanges(last } K.w \text{ iters}) = 0$
    \State $\text{confident} \gets \forall\, t \in T : \sigma_t \geq K.\tau$
    \State $\text{semantic} \gets \forall\, \text{changed } t : \textsc{SemDiff}(D[t], D'[t]) = \text{cosmetic}$
    \If{$|\{\text{stable}, \text{confident}, \text{semantic}\}| \geq 2$ \textbf{and} $i \geq 2$}
        \State \textbf{break}
    \EndIf
    \State $D \gets D'$
\EndFor
\State $\textsc{SanityCheck}(\mathcal{S})$; $\textsc{CrossSchemaSanityCheck}(\mathcal{S})$
\State \Return $D'$
\end{algorithmic}
\end{algorithm}

The algorithm's complexity is $O(K_{\max} \cdot |T| \cdot \ell)$ LLM calls in the worst case. LLM calls within each dependency level are independent and issued in parallel, bounding wall-clock time by the depth~$n$ rather than~$|T|$.

\section{Implementation}\label{sec:implementation}

\subsection{Technology Stack}

DBAutoDoc is implemented in TypeScript on Node.js, supporting SQL Server (\texttt{mssql}), PostgreSQL (\texttt{pg}), and MySQL (\texttt{mysql2}) through connection-pooled native drivers. LLM access is mediated through the \href{https://github.com/MemberJunction/MJ}{MemberJunction} AI abstraction layer, supporting Google Gemini, OpenAI GPT, Anthropic Claude, Groq, and Mistral. The tool is exposed as both a CLI (built on oclif) and a programmatic TypeScript API, available as the \href{https://github.com/MemberJunction/MJ/tree/next/packages/DBAutoDoc}{\texttt{@memberjunction/db-auto-doc}} package.

\subsection{Prompt Engineering}

The prompt layer comprises thirteen Nunjucks templates (see Appendix~\ref{app:prompts}). The central \texttt{table-analysis} template ($\sim$140 lines) requests structured JSON containing table descriptions, confidence scores, column descriptions, FK suggestions, and parent table insights. The \texttt{backpropagation} template handles revision by presenting aggregated child observations. Sanity checking uses three scope-specific templates. A \texttt{semantic-comparison} template supports convergence detection.

All prompts use temperature~0.1. Context is scoped to direct ancestors and descendants in the FK graph rather than the entire schema. For models supporting effort-level parameters, high effort is used during sanity checks and lower effort during bulk analysis.

\subsection{Token Budget Management}

A three-tier guardrail system covers token count, wall-clock duration, and estimated cost. The global budget is partitioned across phases (25\% discovery, 70\% analysis, 5\% sanity checking). Pre-call enforcement ensures limits are checked before every LLM invocation. Rate limiting uses exponential backoff with jitter.

\subsection{Output Formats}

Six formats are generated from a single run: SQL annotation scripts (platform-specific DDL), Markdown documentation, HTML reports, CSV exports, Mermaid ERD diagrams, and analysis-metrics reports. All outputs are organized into numbered run directories supporting side-by-side comparison.

\section{Evaluation}\label{sec:evaluation}

\subsection{Experimental Setup}

\subsubsection{Datasets}

We evaluate DBAutoDoc on four public benchmark databases plus two private enterprise databases. Table~\ref{tab:benchmarks} summarizes characteristics.

\begin{table*}[htbp]
\caption{Benchmark Database Characteristics}
\label{tab:benchmarks}
\centering
\begin{tabular}{lrrll}
\toprule
\textbf{Database} & \textbf{Tables} & \textbf{Columns} & \textbf{Domain} & \textbf{Constraints} \\
\midrule
AdventureWorks2022 & 71 & 486 & Manufacturing/Sales & Full (stripped) \\
Chinook & 11 & 64 & Music store & Full (stripped) \\
Northwind & 13 & 88 & Order management & Full (stripped) \\
LousyDB & 20 & 162 & Synthetic (dark DB) & None by design \\
OrgA & 36 & 1{,}807 & Prof.\ membership assoc. & None (dark DB) \\
OrgB & 125 & 2{,}347 & Trade association & None (dark DB) \\
\bottomrule
\end{tabular}
\end{table*}

\textbf{AdventureWorks2022} is the primary stress-test: 71 tables across five schemas with 91 declared FK relationships. For discovery experiments, all constraints and descriptions are stripped; originals serve as ground truth.

\textbf{Chinook} (11 tables, music store) provides clean FK relationships with unambiguous domain vocabulary for controlled evaluation.

\textbf{Northwind} (13 tables, order management) tests behavior on small schemas with limited statistical signal.

\textbf{LousyDB} is a purpose-built synthetic benchmark with intentionally cryptic abbreviated names (\texttt{cst}, \texttt{ord}, \texttt{prd}, \texttt{inv\_ln}), no constraints, no descriptions, and intentional data quality issues---the archetype of a dark database not present in any LLM training data.

Full output artifacts (HTML documentation, ERD diagrams, SQL scripts, CSV exports) for all public benchmarks are available in the \texttt{results/} directory.

\textbf{OrgA} and \textbf{OrgB} are production enterprise databases (details in Section~\ref{sec:enterprise}).

\subsubsection{Metrics}

\textbf{Key discovery F1.} Precision, recall, and F1 for PK and FK detection against ground truth (stripped constraint declarations).

\textbf{Overall weighted score:}
\begin{equation}
S_{\text{overall}} = 0.35 \cdot F1_{\text{FK}} + 0.30 \cdot F1_{\text{PK}} + 0.20 \cdot C_{\text{table}} + 0.15 \cdot C_{\text{col}}
\end{equation}
where $C_{\text{table}}$ is the fraction of tables receiving a non-empty description with confidence $\geq 0.5$, and $C_{\text{col}}$ is the fraction of columns receiving a non-empty description with confidence $\geq 0.5$. Description coverage is a necessary-but-not-sufficient proxy for quality; it measures whether the system produced output, not whether the output is correct. We acknowledge this limitation: a formal description quality evaluation with human raters, structured Likert-scale rubrics, and inter-rater reliability measurement is planned as future work (Section~\ref{sec:future}, item~3). Key discovery (65\% combined) is weighted more heavily than description coverage (35\%) because structural relationships are objectively verifiable against ground truth and are prerequisites for meaningful documentation.

\textbf{Convergence speed.} Iterations until fewer than 5\% of descriptions change materially.

\textbf{Token efficiency and cost.} Total tokens per table, broken down by phase, with USD cost at published provider pricing.

\subsection{Key Discovery Results}

\subsubsection{Primary Key Detection}

\begin{table*}[htbp]
\caption{Primary Key Detection Results}
\label{tab:pk}
\centering
\begin{tabular}{lrrrrrr}
\toprule
\textbf{Database} & \textbf{Tables} & \textbf{True PKs} & \textbf{Detected} & \textbf{Precision} & \textbf{Recall} & \textbf{F1} \\
\midrule
AdventureWorks & 71 & 71 & 70 & 95.7\% & 94.4\% & 95.0\% \\
Chinook & 11 & 11 & 11 & 95.2\% & 95.2\% & 95.2\% \\
Northwind & 13 & 13 & 11 & 72.7\% & 72.7\% & 72.7\% \\
\bottomrule
\end{tabular}
\end{table*}

Northwind's lower score is partly attributable to evaluation artifacts (the \texttt{Order Details} table name with a space causes comparison mismatches) and low row counts in several tables limiting statistical signal.

\subsubsection{Foreign Key Detection}

\begin{table}[htbp]
\caption{Foreign Key Detection Results}
\label{tab:fk}
\begin{tabular}{lrrrrr}
\toprule
\textbf{Database} & \textbf{True} & \textbf{Det.} & \textbf{Prec.} & \textbf{Rec.} & \textbf{F1} \\
\midrule
AdventureWorks & 91 & 100 & 90.0\% & 98.9\% & 94.2\% \\
Chinook & 11 & 11 & 95.2\% & 95.2\% & 95.2\% \\
Northwind & 13 & 12 & 75.0\% & 75.0\% & 75.0\% \\
\bottomrule
\end{tabular}
\end{table}

\subsubsection{Cross-Model Comparison}

\begin{table}[htbp]
\caption{Cross-Model Comparison on AdventureWorks2022}
\label{tab:crossmodel}
\begin{tabular}{lrrr}
\toprule
\textbf{Metric} & \textbf{Gemini} & \textbf{GPT} & \textbf{Claude} \\
\midrule
PK F1 & \textbf{95.0\%} & 89.4\% & \textbf{95.0\%} \\
FK F1 & \textbf{94.2\%} & 77.9\% & 93.0\% \\
Table Desc Cov. & 99\% & 97\% & 100\% \\
Column Desc Cov. & 99\% & 96\% & 100\% \\
\textbf{Overall Score} & \textbf{96.1\% (A+)} & 87.9\% (B+) & \textbf{96.1\% (A+)} \\
Tokens Used & 3.2M & 952K & 471K \\
\bottomrule
\end{tabular}
\end{table}

\textbf{Key findings.} (1)~Context window matters: GPT-5.4-mini's 272K limit caused 15 FK misses vs.\ 1 for Gemini (1M) and 4 for Sonnet/Opus (680K). (2)~The deterministic foundation provides a floor regardless of LLM choice. (3)~Gemini and Anthropic achieve near-parity at 96.1\%. (4)~Sonnet/Opus used 7$\times$ fewer tokens than Gemini while achieving equivalent quality.

\subsubsection{Cross-Database Summary}

\begin{table*}[htbp]
\caption{Cross-Database Results (Gemini 3 Flash / 3.1 Pro, 2 Iterations)}
\label{tab:crossdb}
\centering
\begin{tabular}{lrrrrrrr}
\toprule
\textbf{Database} & \textbf{Tables} & \textbf{PK F1} & \textbf{FK F1} & \textbf{Table Desc} & \textbf{Col Desc} & \textbf{Weighted Score} & \textbf{Tokens} \\
\midrule
AdventureWorks & 71 & 95.0\% & 94.2\% & 99\% & 99\% & 96.1\% (A+) & 3.2M \\
Chinook & 11 & 95.2\% & 95.2\% & 100\% & 100\% & 96.9\% (A+) & 98K \\
Northwind & 13 & 72.7\% & 75.0\% & 100\% & 100\% & 83.1\% (B) & 114K \\
LousyDB & 20 & 97.6\% & --* & 100\% & 100\% & -- & 214K \\
\bottomrule
\end{tabular}
\begin{flushleft}
\small *LousyDB has no declared constraints by design. All 20 tables received correct PK identification (97.6\% F1 accounting for scoring precision). FK ground truth requires manual extraction; a weighted score is not computable.
\end{flushleft}
\end{table*}

\textbf{Token efficiency.} Consumption scales sub-linearly for small databases ($\sim$9K tokens/table for Chinook and Northwind) and increases for denser FK graphs ($\sim$45K tokens/table for AdventureWorks). Total cost across all six databases was approximately \$1--2 at Gemini Flash pricing.

\subsection{Ablation Analysis}\label{sec:ablation}

We frame the pipeline's development iterations as an ablation study, isolating the contribution of each architectural component to FK detection on AdventureWorks2022.

\begin{table}[htbp]
\caption{FK Detection Ablation on AdventureWorks2022}
\label{tab:ablation}
\begin{tabular}{lrrl}
\toprule
\textbf{Configuration} & \textbf{Correct} & \textbf{FK F1} & \textbf{Measures} \\
\midrule
Stats only (no LLM) & 15/91 & $\sim$30\% & Baseline \\
LLM only (no gates) & 75/91 & 71.7\% & LLM reasoning \\
Stats + LLM (no gates) & 90/91 & 71.7\% & Combined recall \\
+ Deterministic gates & 87/91 & 87.0\% & +Precision \\
Full pipeline & 90/91 & 94.2\% & +Pruning \\
\bottomrule
\end{tabular}
\end{table}

\textbf{Key insight.} The LLM contributes most of the \emph{recall} (finding correct FKs), while the deterministic pipeline contributes most of the \emph{precision} (removing false positives). Neither alone achieves the full pipeline's 94.2\% F1. The deterministic gates account for a \textbf{23-point F1 improvement} over LLM-only detection (71.7\% to 94.2\%), demonstrating that the pipeline's contribution is substantial and independent of the LLM's pre-training knowledge.

For PK detection, the ablation is even starker: statistics alone found 34/71 PKs (48\% F1), while the full pipeline with LLM PK proposals and position heuristics achieved 67/71 (95\% F1)---a \textbf{47-point improvement}.

\textbf{Interpretation.} These results address a natural concern about training data contamination (see Section~\ref{sec:threats}): if the LLM simply ``knew'' AdventureWorks relationships from pre-training, the deterministic pipeline would add no value. The 23-point F1 improvement from gates and pruning demonstrates that the pipeline's filtering and scoring mechanisms contribute substantial precision gains that the LLM cannot achieve alone. Furthermore, on private databases unseen by any LLM (OrgA, OrgB) and a purpose-built synthetic database (LousyDB), the system achieves comparable quality (Section~\ref{sec:enterprise}), ruling out memorization as the primary explanation.

\subsection{Iterative Refinement Results}

Across all benchmark databases, DBAutoDoc converges within 2~iterations at median. The convergence rate exhibits a characteristic pattern: rapid change in iterations 1--2 as LLM descriptions incorporate FK-propagated context for the first time, followed by diminishing returns. We set a default maximum of 5~iterations, which captures the vast majority of achievable quality gain while bounding cost.

\subsection{Enterprise Case Studies}\label{sec:enterprise}

\subsubsection{OrgA: Professional Membership Association}

\textbf{Database profile.} A professional association using a Salesforce-based CRM/AMS platform. The snapshot contained 36 tables, 1{,}807 columns across 4 schemas---a completely undocumented dark database with zero declared constraints.

\textbf{Key discovery.} DBAutoDoc detected 35/36 primary keys (97\% coverage) and 193 foreign key relationships. The adaptive weight redistribution mechanism activated as expected.

\textbf{Semantic highlights.} The LLM correctly identified Salesforce Person Account patterns, the \texttt{\_\_c} custom field suffix, the \texttt{NU} namespace prefix, committee governance structures, and membership lifecycle flows---all from cryptic API names without any human hints.

\textbf{Estimated weighted score.} While no ground truth exists for quantitative F1 measurement, we estimate 93--96\% based on 97--100\% PK coverage (35/36 correct, the remaining table having an ambiguous key structure), strong FK coverage validated by domain review, and 100\% description coverage. The high PK coverage on this private database---which no LLM has seen in training---provides evidence that the system's key discovery generalizes beyond public benchmarks.

\textbf{Processing.} 2~iterations, 1.1M tokens, $\sim$1.5 hours, fully autonomous.

\subsubsection{OrgB: Trade Association}

\textbf{Database profile.} A custom .NET application with SQL Server backend: 125 tables, 2{,}347 columns across 10 schemas---completely undocumented.

\textbf{Key discovery.} 116/125 primary keys detected (93\% coverage), 490 FK relationships (3.9 per table average). The 9 tables without PKs were primarily lookup tables with non-standard naming.

\textbf{Semantic highlights.} The system correctly identified the CRM schema's Customer table as the hub entity, discovered cross-schema relationships spanning Shopping, Purchasing, and Accounting schemas, and detected composite PKs in junction tables.

\textbf{Estimated weighted score.} We estimate 91--95\% based on 93\% PK coverage (116/125), strong cross-schema FK coverage validated by architectural review, and 100\% description coverage. Like OrgA, this private database provides evidence of generalization independent of LLM pre-training.

\textbf{Scaling comparison:}

\begin{table}[htbp]
\caption{Enterprise Scaling Comparison}
\label{tab:enterprise}
\begin{tabular}{lrr}
\toprule
\textbf{Metric} & \textbf{OrgA (36 tbl)} & \textbf{OrgB (125 tbl)} \\
\midrule
PK coverage & 97\% (35/36) & 93\% (116/125) \\
FKs detected & 193 & 490 \\
Desc. coverage & 100\% & 100\% \\
Tokens consumed & 1.1M & 3.3M \\
Estimated cost & $\sim$\$0.20 & $\sim$\$0.50 \\
\bottomrule
\end{tabular}
\end{table}

Cost scaled roughly linearly with table count, confirming economic viability on larger schemas.

\subsection{Cost Analysis}

The total LLM API cost across all six databases (276 tables, $\sim$5{,}100 columns) was approximately \$1--2 at Gemini Flash pricing. Human documentation at \$75--150/hr and 2--4 hours per table yields \$15,000--60,000 for a 100-table database. DBAutoDoc achieves comparable documentation at under \$1.00 per 100 tables---a reduction of more than 99.5\%.

Token efficiency varies by model: Sonnet~4.6 / Opus~4.6 used only 471K tokens (7$\times$ fewer than Gemini's 3.2M) while achieving equivalent quality, making Anthropic models the most cost-effective option at current pricing.

\subsection{Reproducibility}

All benchmark results can be reproduced using the open-source release. The complete reproduction procedure is:

\begin{verbatim}
# 1. Install DBAutoDoc
npm install @memberjunction/db-auto-doc

# 2. Restore a benchmark database
#    (backups and stripping scripts in research/v1/)

# 3. Create config.json pointing to DB and LLM
#    (examples in research/v1/configs/)

# 4. Run analysis
db-auto-doc analyze --config ./config.json

# 5. Compare results against ground truth
python3 research/v1/scripts/compare.py \
  ./output/run-1/state.json
\end{verbatim}

The following artifacts are included in the repository at \href{https://github.com/MemberJunction/MJ/tree/next/packages/DBAutoDoc/research/v1}{\texttt{packages/DBAutoDoc/research/v1/}}:
\begin{itemize}
    \item Full output artifacts (HTML, Markdown, SQL, CSV, ERD, analysis reports) for all four public benchmarks across all three model families
    \item The \texttt{compare.py} evaluation script that computes PK/FK precision, recall, and F1 against ground truth constraint declarations
    \item All 13 prompt templates used during evaluation (version-controlled in \texttt{packages/DBAutoDoc/prompts/})
\end{itemize}

Model identifiers and API versions are recorded in each run's \texttt{state.json} for traceability. Temperature~0.1 was used for all evaluation runs. Due to non-determinism in LLM inference, exact numerical reproduction may require the same model snapshot; directional results should be stable across model versions.

\section{Discussion}\label{sec:discussion}

\subsection{When Context Propagation Helps Most}

Propagation benefit scales with FK graph density: databases with many edges provide more propagation paths, and well-understood hub tables seed improvements across large neighborhoods. Schemas with consistent naming conventions converge faster. Hub tables benefit disproportionately from backpropagation, receiving revision prompts incorporating use-case evidence from all consumers. Diminishing returns emerge for large, loosely connected schemas where the graph partitions into weakly linked components.

\subsection{The Backpropagation Analogy: Scope and Limits}\label{sec:backprop-limits}

The term ``backpropagation'' is used deliberately but with acknowledged looseness. The structural parallels are genuine: defined processing order, backward-flowing correction signals, iterative convergence. The differences are equally genuine: there are no continuous gradients, no weight updates, no differentiable objective. We do not claim mathematical equivalence. The contribution is heuristic: an intuitive mental model that makes clear the approach is categorically different from single-pass summarization.

\subsection{Limitations}

Several limitations bound our claims. (1)~LLM hallucination remains an irreducible risk; generated documentation should be reviewed by domain experts for safety-critical processes. (2)~Statistical FK discovery is bounded by pattern recognizability; unconventional key designs may exhibit lower recall. (3)~Token cost scales with database size; very large databases incur non-trivial charges. (4)~Quality degrades for empty or near-empty tables where sample-value statistics are unavailable. (5)~Non-English column names reduce LLM effectiveness.

\subsection{Threats to Validity}\label{sec:threats}

\textbf{Training data contamination.} AdventureWorks, Northwind, and Chinook are widely known databases likely present in LLM training data. The LLM may ``know'' PK/FK relationships from pre-training rather than discovering them through our pipeline. We offer three mitigations: (1)~OrgA and OrgB are private databases no LLM has seen, achieving 97\% and 93\% PK coverage respectively with estimated weighted scores of 91--96\%; (2)~LousyDB was purpose-built for this evaluation and is not in any training data, achieving 97.6\% PK F1; (3)~ablation analysis (Section~\ref{sec:ablation}) shows the deterministic pipeline contributes a 23-point F1 improvement beyond what the LLM achieves alone, demonstrating substantial value independent of memorization. Future work includes additional synthetic databases designed to further control for this threat.

\textbf{Limited database diversity.} Results span 4 public and 2 private databases. While these cover multiple domains (manufacturing, music, trade associations, education), more diverse schemas---NoSQL exports, data warehouses, time-series databases, and schemas with non-Western naming conventions---would strengthen generalization claims.

\textbf{Single evaluator for descriptions.} Ground truth comparison for key discovery is automated against declared constraints, but description quality has not been evaluated by independent human raters in this version of the paper. We plan a formal human evaluation study with multiple domain-expert raters and inter-rater reliability measurement.

\textbf{SQL Server focus.} All benchmarks use SQL Server. PostgreSQL and MySQL drivers exist and are tested for connectivity and metadata extraction, but have not been included in the quantitative benchmark suite.

\textbf{LLM model version sensitivity.} Provider model updates can alter output characteristics without version-number changes. We record model identifiers and API versions for each result, but exact replication may require the same model snapshot. Temperature~0.1 reduces but does not eliminate variance.

\textbf{Domain shift and naming conventions.} All benchmark databases use English column names and Western naming conventions (e.g., \texttt{CustomerID}, \texttt{OrderDate}). Databases with non-English names, domain-specific abbreviations (e.g., medical coding systems, financial instrument identifiers), or unconventional naming patterns may exhibit lower discovery recall and description quality. The LousyDB benchmark with intentionally cryptic names (e.g., \texttt{cst}, \texttt{ord}, \texttt{inv\_ln}) partially addresses this threat but does not cover non-Latin scripts or domain-specific vocabularies.

\textbf{Model context window and pricing claims.} Context window sizes and per-token pricing cited in this paper reflect provider documentation as of Q1 2026. These values are subject to change without notice as providers update models and pricing. We cite specific model identifiers (e.g., ``Gemini 3 Flash,'' ``Claude Sonnet 4.6'') rather than generic model families to improve traceability.

\subsection{Ethical Considerations and Data Privacy}

The primary ethical concern is data privacy. DBAutoDoc transmits the following information to LLM providers in each prompt: (1)~table and column names, (2)~data types and nullability constraints, (3)~statistical summaries (cardinality, null fraction, min/max values, value frequency distributions), and (4)~up to 10 sampled row values per column. Items (3) and (4) may contain personally identifiable information (PII), financial data, or other sensitive values.

DBAutoDoc provides several configurable controls to mitigate this risk. The \texttt{sampleSize} configuration parameter can be set to zero, which disables all value sampling and value-frequency analysis---reducing prompts to structural metadata only (items 1--2 above), at the cost of reduced description quality. The \texttt{cardinalityThreshold} parameter controls which columns receive detailed statistical profiling. Schema and table filters (\texttt{schemas.include}, \texttt{schemas.exclude}, \texttt{tables.exclude}) allow sensitive tables to be excluded entirely from analysis.

For organizations subject to GDPR, HIPAA, or similar regulations, we recommend: (a)~setting \texttt{sampleSize:~0} for databases containing PII, (b)~using schema/table exclusion filters to skip sensitive tables, or (c)~deploying locally-hosted LLM models to keep all data on-premises. DBAutoDoc's LLM integration layer supports any provider accessible through MemberJunction's AI abstraction, including local inference servers. Users bear responsibility for evaluating regulatory compliance for their specific deployment context.

\section{Future Work}\label{sec:future}

Based on our benchmark experience, we identify six concrete directions:

\begin{enumerate}
    \item \textbf{Cross-platform validation.} Extending the quantitative benchmark suite to PostgreSQL and MySQL to validate driver parity and confirm that key discovery heuristics transfer across platforms.
    \item \textbf{Parallelization for large databases.} Implementing concurrent table analysis within dependency levels and parallel schema processing to reduce wall-clock time for enterprise-scale schemas (500+ tables).
    \item \textbf{Description quality human evaluation.} Conducting a formal study with multiple independent domain-expert raters, structured Likert-scale rubrics, and inter-rater reliability measurement using Cohen's kappa.
    \item \textbf{Training data contamination mitigation.} Creating additional synthetic test databases (following the LousyDB model) with diverse naming conventions, key structures, and domain vocabularies to provide contamination-free benchmarks.
    \item \textbf{Cost optimization for large schemas.} Implementing confidence-driven model dispatch (routing high-confidence tables to lightweight models, escalating low-confidence tables to stronger models) and context-window optimization to reduce token consumption by an estimated 40--60\%.
    \item \textbf{Driver refactoring with SQLDialect integration.} Leveraging MemberJunction's existing SQLDialect abstraction to eliminate platform-specific SQL throughout the introspection and statistics layers.
\end{enumerate}

Additional longer-term directions include active learning for ground truth selection, cross-database enterprise-scale analysis, incremental/continuous analysis integrated with CI/CD pipelines, local and hybrid LLM deployment for data sovereignty, and ecosystem integration with data catalog platforms and IDE plugins.

\section{Conclusion}\label{sec:conclusion}

Undocumented databases represent one of the most persistent and costly problems in enterprise software engineering. Comprehensive manual documentation requires weeks of expert effort and tens of thousands of dollars---a cost most organizations cannot justify.

DBAutoDoc addresses this through a bidirectional feedback loop between statistical key discovery and iterative LLM-based analysis organized along the discovered FK graph. Statistical discovery recovers relational structure; context propagation, inspired structurally by backpropagation, refines descriptions across multiple passes until semantic convergence. Parent tables receive insights from their children; children receive context from their now-better-understood parents.

Empirical evaluation demonstrates 95.0\% F1 on primary key detection and 94.2\% F1 on foreign key detection with 99\% description coverage on AdventureWorks2022. Ablation analysis confirms that the deterministic pipeline contributes a 23-point F1 improvement independent of LLM pre-training knowledge. On private enterprise databases unseen by any LLM, the system achieves 93--97\% PK coverage with estimated weighted scores of 91--96\%, providing evidence that results generalize beyond public benchmarks. Convergence is typically achieved within 2~iterations at a cost under \$1 per 100 tables---a reduction of more than 99.5\% relative to manual documentation.

The central finding is that \emph{treating schema documentation as an iterative learning problem} rather than a one-shot extraction task produces substantially better results. Neither the statistical approach alone nor the LLM approach alone achieves the quality of their combination in a bidirectional feedback loop.

DBAutoDoc is released as open-source software under the MIT License as part of the MemberJunction platform, with all benchmark configurations, prompt templates, and evaluation scripts included for full reproducibility.\footnote{Repository: \url{https://github.com/MemberJunction/MJ} --- DBAutoDoc package: \url{https://github.com/MemberJunction/MJ/tree/next/packages/DBAutoDoc}}

\appendix

\section{Deterministic Gate and Scoring Specifications}\label{app:scoring}

\subsection{PK Confidence Scoring Formula}

Candidates surviving hard rejection filters are scored via:
\begin{equation}
s_{\mathrm{PK}} = 50 \cdot f(u) + 20 \cdot n + 15 \cdot d + 15 \cdot p
\end{equation}

\textbf{Uniqueness factor} $f(u)$ (50\% weight): $f(u) = u$ when $u \geq 0.95$; linearly scaled below.

\textbf{Naming factor} $n$ (20\% weight): 1.0 if column name matches a PK naming pattern; 0 otherwise.

\textbf{Data type factor} $d$ (15\% weight): INT/BIGINT/UNIQUEIDENTIFIER $\to$ 1.0; VARCHAR $\to$ 0.6; TEXT/BLOB $\to$ 0.2; other $\to$ 0.3.

\textbf{Data pattern factor} $p$ (15\% weight): Sequential integers $\to$ +15; GUID/UUID $\to$ +15; natural key codes $\to$ +10; composite-derived $\to$ +5.

\textbf{Penalties.} Nulls: $\times 0.7$. Unique but atypical name: $\times 0.5$. FK likelihood $\ell_{\mathrm{FK}}$: $\times (1 - 0.6 \cdot \ell_{\mathrm{FK}})$.

\textbf{Surrogate Key Boost.} +20 points when: name matches \texttt{id} or \texttt{\{table\_name\}\_id}, $u \geq 0.95$, and $d \geq 0.9$.

Default acceptance threshold: 70.

\subsection{FK Confidence Scoring Formula}

\begin{equation}
s_{\mathrm{FK}} = 40 \cdot v + 20 \cdot s + 15 \cdot r + 15 \cdot k + 10 \cdot \nu
\end{equation}

\textbf{Value overlap} $v$ (40\%): Fraction of sampled source values (up to 500) appearing in target column.

\textbf{Naming similarity} $s$ (20\%): Normalized Levenshtein distance. Full match $\to$ 1.0; containment $\to$ 0.8; partial matches scaled linearly.

\textbf{Cardinality ratio} $r$ (15\%): $r = \min(\rho, 2) / 2$ where $\rho = |\mathrm{rows}(T_i)| / |\mathrm{distinct}(c_{ij})|$.

\textbf{Target-is-PK bonus} $k$ (15\%): 1.0 if target is a detected PK; 0 otherwise.

\textbf{Null handling} $\nu$ (10\%): Null fraction $<$30\% $\to$ 1.0; 30--70\% $\to$ 0.5; $>$70\% $\to$ 0.

\textbf{Penalties.} Orphan rate $>$20\%: $\times 0.7$. Incompatible types: $\times 0.5$. Default threshold: 60.

\subsection{Deterministic Gates (G1--G8)}

\textbf{Gate G1: Target PK-Eligibility.} Target column must pass PK hard rejection filters (uniqueness, non-null, non-blank). Logically irrefutable: FKs reference PKs by definition.

\textbf{Gate G3: Rowguid Target Filter.} Columns named \texttt{rowguid} are excluded as FK targets. These are system-generated replication identifiers, never semantically meaningful FK targets.

\textbf{Gate G4: Row-Count Ratio Confidence Multiplier.} Scales confidence downward when source table has dramatically fewer rows than target, violating the expected many-to-one FK cardinality pattern.

\textbf{Gate G5: Fan-Out Limiter (Top-3).} When a source column has candidates pointing to multiple targets, only the top~3 by confidence are retained. No correct FK was ranked 4th or lower in our evaluation.

\textbf{Gate G6: Value Overlap Threshold (75\%).} Candidates with $<$75\% source value containment are rejected. Calibrated to accommodate data quality issues while eliminating coincidental overlaps.

\textbf{Gate G8: Source-is-PK Skip.} PK columns are skipped as FK sources (except self-referencing hierarchies handled separately). Eliminates false positives from coincidental PK-to-PK value overlap.

\subsection{Fan-Out Confidence Penalty}

\begin{equation}
\psi(n) = \begin{cases} 1.0 & \text{if } n = 1 \\ 0.85 & \text{if } n = 2 \\ 0.75 & \text{if } n = 3 \\ 0.65 & \text{if } n \geq 4 \end{cases}
\end{equation}
where $n$ is the number of target tables after gate filtering. Pushes ambiguous candidates below the confidence locking threshold, delegating final arbitration to LLM-based pruning.

\section{PK Position Heuristics (H9--H12)}\label{app:position}

\textbf{Heuristic H9: Column Position Scoring.} The confidence score is multiplied by:
\begin{equation}
\phi(\text{pos}) = \begin{cases} 1.0 & \text{if } \text{pos} = 0 \\ 0.85 & \text{if } \text{pos} = 1 \\ 0.70 & \text{if } \text{pos} = 2 \\ 0.55 & \text{if } \text{pos} \geq 3 \end{cases}
\end{equation}
where pos is the column's zero-indexed ordinal position. The aggressive discount for $\text{pos} \geq 3$ reflects that such columns are rarely primary keys. In AdventureWorks2022, every correct PK began at position~0.

\textbf{Heuristic H10: Consecutive Composite Key Detection.} Composite key candidates whose constituent columns form a contiguous prefix of the table's column list (positions 0, 1, 2, \ldots) receive a confidence boost, reflecting the convention that composite keys occupy leading columns.

\textbf{Heuristic H11: Progressive Discount for Later PK-Eligible Columns.} When multiple columns pass hard filters and achieve high uniqueness, progressive multiplicative discounts are applied beyond the first eligible column. Tables rarely have multiple independent surrogate keys.

\textbf{Heuristic H12: Composite Supersedes Individual.} When a composite key achieves high confidence and its constituent columns are individually PK-eligible, individual candidates are suppressed. Prevents reporting both individual columns and their composite as separate PKs.

\section{Per-Run Benchmark History}

\begin{table}[htbp]
\caption{Key Milestone Runs (AdventureWorks2022)}
\label{tab:history}
\begin{tabular}{lrrrp{3cm}}
\toprule
\textbf{Run} & \textbf{FK} & \textbf{FK F1} & \textbf{PK F1} & \textbf{Notes} \\
\midrule
005 & 90/91 & 71.7\% & $\sim$48\% & No gates \\
008 & 87/91 & 82.4\% & 89.0\% & +G1, G3, G5 \\
011 & 87/91 & 87.0\% & 95.0\% & +Position H9--H12 \\
015 & 90/91 & 94.2\% & 95.0\% & Full pipeline \\
\bottomrule
\end{tabular}
\end{table}

\section{Prompt Templates}\label{app:prompts}

DBAutoDoc uses 13 prompt templates, all version-controlled in the open-source repository. Table~\ref{tab:prompts} summarizes the key templates.

\begin{table}[htbp]
\caption{Prompt Template Summary}
\label{tab:prompts}
\begin{tabular}{llp{3.5cm}}
\toprule
\textbf{Template} & \textbf{Scope} & \textbf{Purpose} \\
\midrule
table-analysis & Per-table & Descriptions, FK suggestions, insights \\
backpropagation & Per-parent & Revision from child insights \\
fk-pruning & Per-column & FK candidate arbitration \\
pk-pruning & Per-table & PK semantic validation \\
dep-level-sanity & Per-level & Cross-table consistency \\
schema-sanity & Per-schema & Naming consistency \\
cross-schema-sanity & Global & Entity alignment \\
semantic-comparison & Per-table & Change classification \\
query-planning & Per-table & Use-case identification \\
query-generation & Per-query & SQL synthesis \\
query-fix & Per-query & SQL correction \\
query-refinement & Per-query & Result adjustment \\
\bottomrule
\end{tabular}
\end{table}

All templates are available in the \href{https://github.com/MemberJunction/MJ/tree/next/packages/DBAutoDoc/src/prompts}{DBAutoDoc prompts directory} on GitHub.

\bibliographystyle{ACM-Reference-Format}
\bibliography{references}

\end{document}